\documentstyle[12pt,epsf,epsfig]{article}

\def\be{\begin{equation}}
\def\bea{\begin{eqnarray}}
\def\ee{\end{equation}}
\def\eea{\end{eqnarray}}
\def\to{\rightarrow}
\def\ov{\overline}
\def\ra{\rangle}
\def\la{\langle}
\def\r{\right}
\def\l{\left}

\def\a{\alpha}
\def\b{\beta}

\def\d{\delta}
\def\C{{\cal C}}
\def\D{\Delta}

\def\th{\theta}
\def\s{\sigma}
\def\S{{\bf \sigma}}
\def\ti{\hat}
\def\pr{\prime}
\def\p{\partial}
\def\B{\tilde{B}}
\def\c{\tilde{c}}

\begin{document}

%\title{The Replica Symmetric Approximation for Random Heteropolymers
%is not equivalent to the Annealed Approximation}

\title{Exactness of the Annealed and the Replica Symmetric Approximations
for Random Heteropolymers}
\author{Ugo Bastolla$^{(1,2)}$ and Peter Grassberger$^{(1)}$\\
{\sl \small $^{(1)}$HLRZ, Forschungszentrum J\"ulich, D-52425 J\"ulich,
Germany}\\
{\sl \small
$^{(2)}$Max Planck Institute for Colloids and Interfaces, D-14424 Potsdam,
Germany}
}
\maketitle
 
\medskip
\medskip

\begin{abstract}
We study a heteropolymer model with random contact interactions
introduced some time ago as a simplified model for proteins. The model
consists of self-avoiding walks on the simple cubic lattice,
with contact interactions between nearest neighbor pairs. For
each pair, the interaction energy is an independent Gaussian variable
with mean value $B$ and variance $\D^2$. For this model
the annealed approximation is expected to become exact for low disorder,
at sufficiently
high dimension and in the thermodynamic limit. We show that corrections
to the annealed approximation in the 3-d high temperature phase are
small, but do not vanish in the thermodynamic limit, and are in
good agreement with our replica symmetric calculations. Such corrections
derive from the fact that the overlap between two typical chains is nonzero.
We explain why previous authors had come to the
opposite conclusion, and discuss consequences for the thermodynamics
of the model.
Numerical results were obtained by simulating chains  of 
length $N\leq 1400$ by means of the recent PERM algorithm,
in the coil  and molten globular phases, well above the freezing temperature.
 
\end{abstract}

\section{Introduction}

Apart from their extreme biological importance, proteins are also very
interesting objects from the point of view of statistical
mechanics. They possess a very well defined native structure which they
are able to find in a short time among a potentially huge number of
competing ones, and in spite of many metastable states. How proteins
reconcile the stability of the native
structure with the requirement that this structure is rapidly reached
constitutes the essence of the fascinating  and still open
protein folding problem \cite{PF}
 
An interesting question is whether the property of folding is a generic
property of randomly assembled polypeptidic chains, regardless of
their biological function, or is a special property that has
evolved through natural selection.
This kind of question makes
the protein folding problem a bridge between theoretical biology and
the statistical mechanics of disordered systems.
Motivated by this, numerous authors have studied simple models of random 
heteropolymers 
\cite{BW,GO,SG,SK,SGcube,SG2,SG3,CT,SGdes,IMP,KK,PWW,PGJKT,PGJT,CS,KT,MSVO},
see \cite{GOP} for a review.
 
In the following, we shall discuss only the `random bond model' introduced
independently by Garel and Orland \cite{GO} and by Shakhnovich and Gutin
\cite{SG}. More precisely, in our numerical simulations we will study a
lattice version of this model. Preliminary results of this work have already
been presented in \cite{proc}.
A ``protein'' with $N+1$ ``amino acids''
is represented as a self-avoiding walk \cite{DeGennes} of $N$ steps on the
simple cubic lattice. Each pair $(i,j)$ of non-bonded monomers on nearest
neighbor lattice sites contributes to the total energy an amount given by 
an independent and identically distributed (iid) Gaussian variable 
$B_{ij}$ with mean $B'$ and variance $\D'^2$. Formally, one defines
the contact map of configuration $\C$, $\S(\C)$, as the matrix of
binary variables $\s_{ij}\in\{0,1\}$, with $i,j\in \{0,\cdots N\}$,
such that
\be
   \s_{ij}(\C)=\left
	\{\begin{array}{ll} 1 & \hbox{if $i$ and $j$ are in contact
                                         and non-bonded } \\
                                    0 & \hbox{otherwise.}
           \end{array}\right. \label{contact}
\ee
The energy of the model can then be written as
\be
   E(\C,\{B\})=\sum_{i<j} \s_{ij}(\C)B_{ij},
\ee
with $\ov{B_{ij}} = B', \: \ov{B_{ij}^2}-\ov{B_{ij}}^2 = \D'^2$. For a given
realization of the interaction energies $B_{ij}$ (representing a protein
sequence in the biological analogy), the partition sum $Z_N$ at
temperature $T$ can be formally computed as
\be
   Z_N\{B_{ij}\} = \sum_{\cal C} e^{-E(\C,\{B\})/k_BT}\;,
\ee
where the sum over configurations ${\cal C}$ runs over all self-avoiding 
$N$-step walks. Obviously, the above expression depends only on the
variables $\D=\D'/k_BT$ and $B=B'/k_BT$, i.e. we have a
two-parameter phase diagram in the variables $B$ and $\D$. The main advantage
in using $B$ as one of
the independent variables instead of $T$ or $\beta=1/k_BT$ is that we can pass
continuously from positive (repulsive, hydrophilic) to negative (hydrophobic)
$B$.

As usual with random models, we have to evaluate the quenched average of
the free energy. This is a very difficult task, while it is rather easy to
perform an annealed average over the disorder. For several models of random
spin systems it is well known that such an
annealed approximation becomes exact in the high temperature phase, in the 
thermodynamic limit and at sufficiently large dimension. The same is thought
to be true for the present model. It was indeed predicted in \cite{SG} that 
the annealed approximation 
becomes exact in 3 dimensions when the chain length tends to infinity.
For this to be true it is necessary that the overlap between two randomly 
chosen replicas
with the same realization of disorder vanishes in the limit $N\to\infty$.

Numerical tests of this prediction have been made in the past for chains 
of length $\leq 36$, mostly by means of exact enumerations of maximally 
compact chains of length 27 \cite{PGJKT,PGJT}. These authors found deviations 
(replica overlap is non-zero) which seemed to decrease with $N$. A similar 
result even for $d=2$ was found in \cite{MSVO}, where also Monte Carlo 
simulations of very short chains were used (up to $N=64$).
But it is clear that tests with such short
chains can hardly be significant. In the present paper we shall present 
Monte Carlo simulations for chains of length up to $N=1400$. These 
simulations are made with the PERM algorithm developed recently by one of 
us \cite{PERM}, and applied successfully to a number of different polymer 
problems \cite{stiff,prot,unmix,OSAW}.

We study the corrections to the annealed approximation using two
different approaches. First, we compute them using the
replica method and assuming replica symmetry, which is believed to hold
for low disorder. Even if a full computation was not possible,
the expected behavior was well confirmed by numerical simulations. Second,
we notice that
corrections to the annealed approximation in the weak disorder limit 
can be related exactly to the average overlap between pairs of homopolymers 
(without any disorder). We give strong theoretical arguments that this 
overlap does not vanish in the limit $N\to\infty$. We also calculate it by 
means of Monte Carlo simulations. Unlike in the previous case, these
simulations do not involve the averaging over the disorder and thus can
be applied to larger systems.

The two methods agree with each other and show that the corrections to the
annealed  approximation are small in $d=3$, but do not vanish in the
thermodynamic  limit.
%More precisely, for not too strong disorder one has a coil-globule transition
%which has the same  qualitative features as the 
%$\theta$-transition in bad solvents.
Deviations from the annealed 
approximation are larger in the coil (high-temperature) phase and very
small in the collapsed (globular) phase.

The annealed approximation is presented in sec.2 and compared to results 
of Monte Carlo simulations. In order to explain the observed deviations, 
we study in sec.3 a scenario where the overlap is non-zero 
but replica symmetry is unbroken. We again compare theoretical predictions
with simulation results. The relationship between the weak disorder limit 
and homopolymer overlap is discussed in sec.4. Additional thermodynamic 
considerations are presented in sec.5, and our final conclusions are 
drawn in sec.6. The PERM algorithm used for the simulations is discussed in 
an appendix.

\section{Annealed Approximation}

In thermodynamic systems with quenched disorder we have to consider the
average of the free energy per monomer over individual
realizations of disorder $\{B_{ij}\}$ which formally is given by

\be
   F_N(B,\D) = -{1\over \b N} \;\overline{\log\l(Z_N\{B_{ij}\}\r))} \equiv 
    -{1\over \b N} \;\prod_{i<j} \int dB_{ij} {e^{-(B_{ij}-B)^2/2\D^2}\over
     \D\sqrt{2\pi}}   Z_N\{B_{ij}\} \;.
\ee

As for most random systems, this cannot be evaluated in closed form.
Much easier to evaluate is the annealed approximation 
\be
   F_{N, \rm ann}(B,\D) = -{1\over N} \;\log {\overline Z_N}
                                  \label{annealed}
\ee
obtained by taking the disorder average before taking the log. 
Here the Gaussian integrals can be done explicitly, with the result 
\be
   {\overline Z_N} = \sum_{\C} e^{-(B-{1\over 2}\D^2)\sum_{i<j} \s_{ij}(\C)}\;.
\ee
Since this is the partition sum for a homopolymer with pair energy 
\be
   \B= B-{1\over 2}\D^2, \label{ann2}
\ee
we see that \cite{SG}
\be
   F_{N, \rm ann}(B,\D) = F_N(\B,0)\;.    \label{F-annealed}
\ee

Therefore, all thermodynamic variables can be expressed in the annealed 
approximation in terms of an equivalent homopolymer with shifted 
interaction strength.  This relationship is easiest for those observables 
whose definition does not involve a derivative with respect to temperature,
such as the gyration and end-to-end radii, and the density of 
non-bonded nearest neighbor contacts $c$. The latter is defined as the 
average number of nn. contacts between non-consecutive monomers divided 
by $N$. For these observables, we have 
\be
   R_{N, \rm ann}(B,\D) = R_N(\B,0)    \label{R-ann}
\ee
and 
\be
   c_{\rm ann}(B,\D) = c(\B,0)\equiv \c,    \label{c-ann}
\ee
precisely as in eq.(\ref{F-annealed}).

For energy $U$ and entropy $S$ the relations are less simple, since these 
involve derivatives of the free energy with respect to $T$ which are
changed into derivatives with respect to $B$ and $\D$ by our
convention of using $T=1$. For the energy per monomer it holds
\be
   U_{N, \rm ann}(B,\D) = {B- \D^2 \over \B } U_N(\B,0)=
\l({B- \D^2 \over \B }\r)\c \;,  \label{U-anneal}
\ee
where we used the fact that the energy for homopolymers is $U_N(\B,0) = \c\B$.
For the specific entropy
$S_N(B,\D) = -{\partial \over\partial T} F_N(B,\D,T)|_{T=1}$ 
we use $F_N(B,\D,T) = \\ 
T F_N(B/T,\D/T,1)$ together with eq.(\ref{F-annealed}), 
and obtain 
\be
   S_{N, \rm ann}(B,\D)=S_N(\B,0) - {\D^2\over 2\B} U_N(\B,0)\;. \label{S-anneal}
\ee

The number of configurations
with fixed $c$ should increase as $\exp(Nf(c))$ for large $N$, i.e. $f(c)$ is
the entropy density in the fixed-$N$, fixed-$c$ ensemble. For homopolymers, the
ensemble with fixed $\B$ becomes equivalent to the fixed-$c$ ensemble in
the limit $N\to\infty$, Thus $c$ becomes a non-fluctuating function of $\B$,
$c = c(\B)\equiv \c$, and the above formula becomes simply
\be
   S_{N, \rm ann}(B,\D)= f(\c)-{\D^2\over 2}\c\qquad {\rm for} N\to\infty\;,
\ee
where $c(\ti{B})$ is the solution of the saddle point equation
\be
   f'(\c) \equiv {\p f(c)\over \p c}|_{c=\c} = \B .
                                     \label{fprim}
\ee

The condition for thermodynamic stability is that the second derivative 
of $f$ should be negative, corresponding to $F$ being minimal.
This is equivalent to requiring that the specific heat is positive.
In fact, the specific heat for a homopolymer is given by

\be
C_V=B{\p c\over\p T}=-B^2\l({\p^2f\over\p c^2}\r)^{-1}, \label{Cv}
\ee
which has been obtained by
deriving both sides of Eq.(\ref{fprim}) with respect to $T$.

Homopolymers with attraction between unbonded nearest neighbors show a collapse
(``theta'') transition where the specific heat diverges. Thus we expect that
the second derivative $\p^2f/\p c^2$ vanishes at the theta-point $c=c_\th$
(the precise value of the transition point depends on the lattice considered).

\begin{figure}[ht]
  \centerline{
    \psfig{file=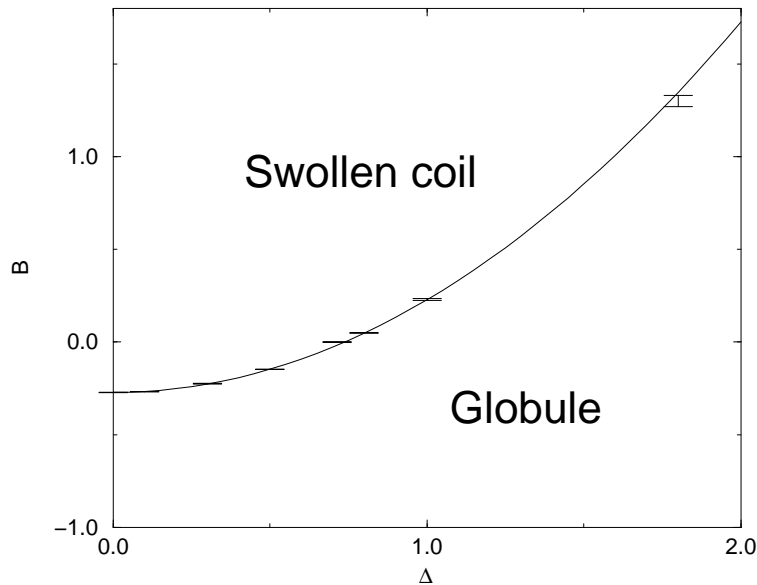,width=10cm}}
\caption{Collapse transition line. The solid line is the annealed
    prediction, $B-\D^2/2=B_\th=-0.269$. Numerical data are obtained by
    measuring the end-to-end distance. At $\D\geq 2.5$ (not shown)
    there are significant deviations from the annealed approximation,
    that are most likely due to a direct freezing from the swollen phase.}
\label{fig:theta}
\end{figure}

The annealed approximation is supposed to be valid both above and below the
theta transition. 
At very low temperatures and very large disorder, it has to break 
down since otherwise the entropy would become negative, according to
eq.(\ref{S-anneal}). This signals another phase transition, the so-called 
freezing transition. We shall not discuss this regime in this paper, but will 
treat it in a forthcoming publication.

Since the theta point is a tricritical point \cite{DeGennes,PERM}, its 
upper critical dimension is d=3. Therefore, we expect that in three dimensions
the ``swelling factor'' is constant,
\be
   \langle R^2\rangle/ N \approx const    \label{R-theta}
\ee
at the theta point, up to logarithmic corrections \cite{dupla,PERM,HS}. Here, 
$R$ is any measure of the size of the polymer, such as the end-to-end 
distance or 
the gyration radius. We expect that this is still true for heteropolymers, 
as long as we are not yet in the frozen regime.
While eq.(\ref{R-theta}) gives the most precise numerical estimate of 
the collapse transition (with $B_\th = -0.2690\pm 0.0002$ \cite{PERM}), 
estimates with similar precision can be obtained from the convexity 
of the free energy \cite{unmix}, and the volume dependence of the free 
energy in case of periodic boundary conditions \cite{stiff}.

\begin{figure}[ht]
  \centerline{
    \psfig{file=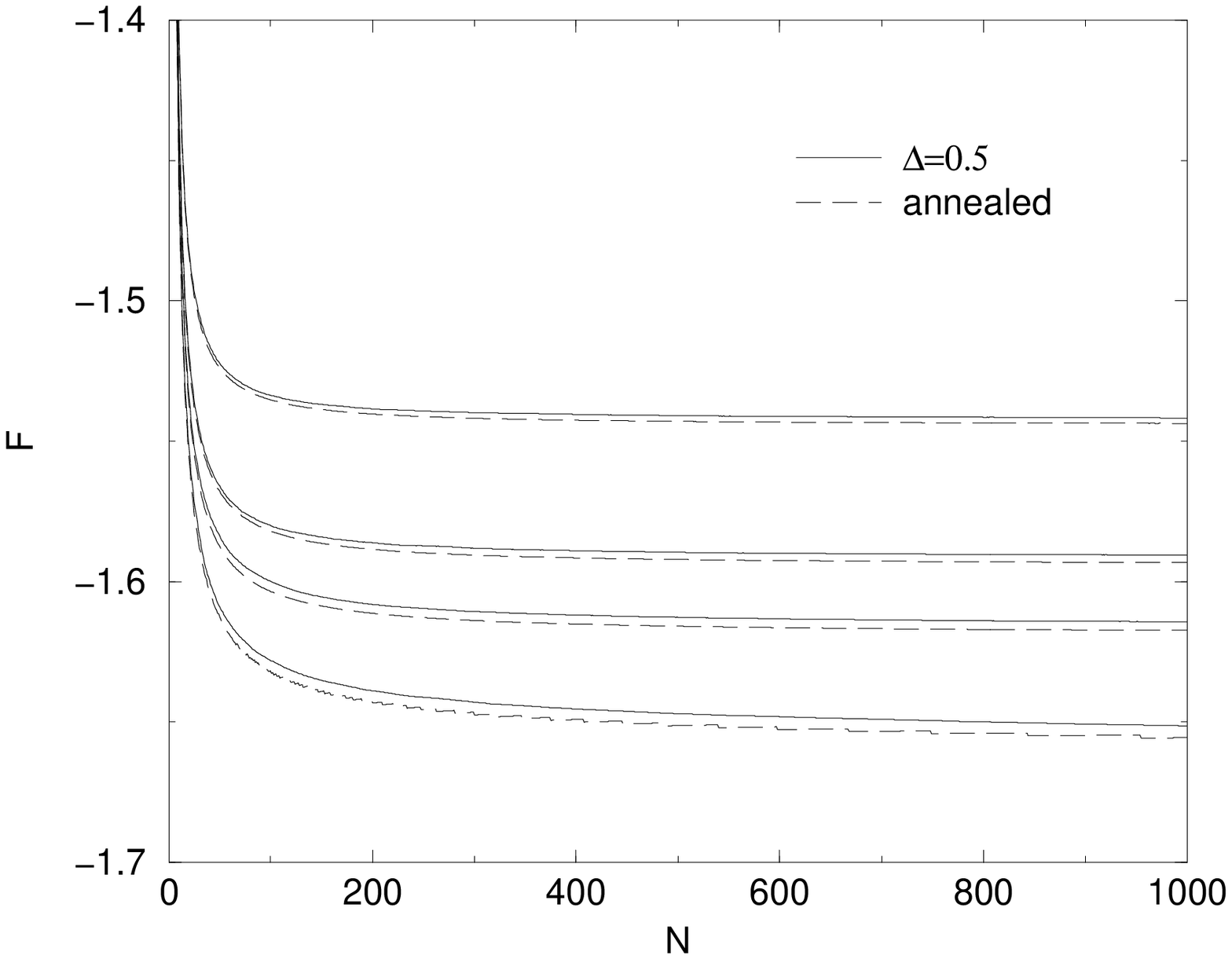,width=7cm}
    \psfig{file=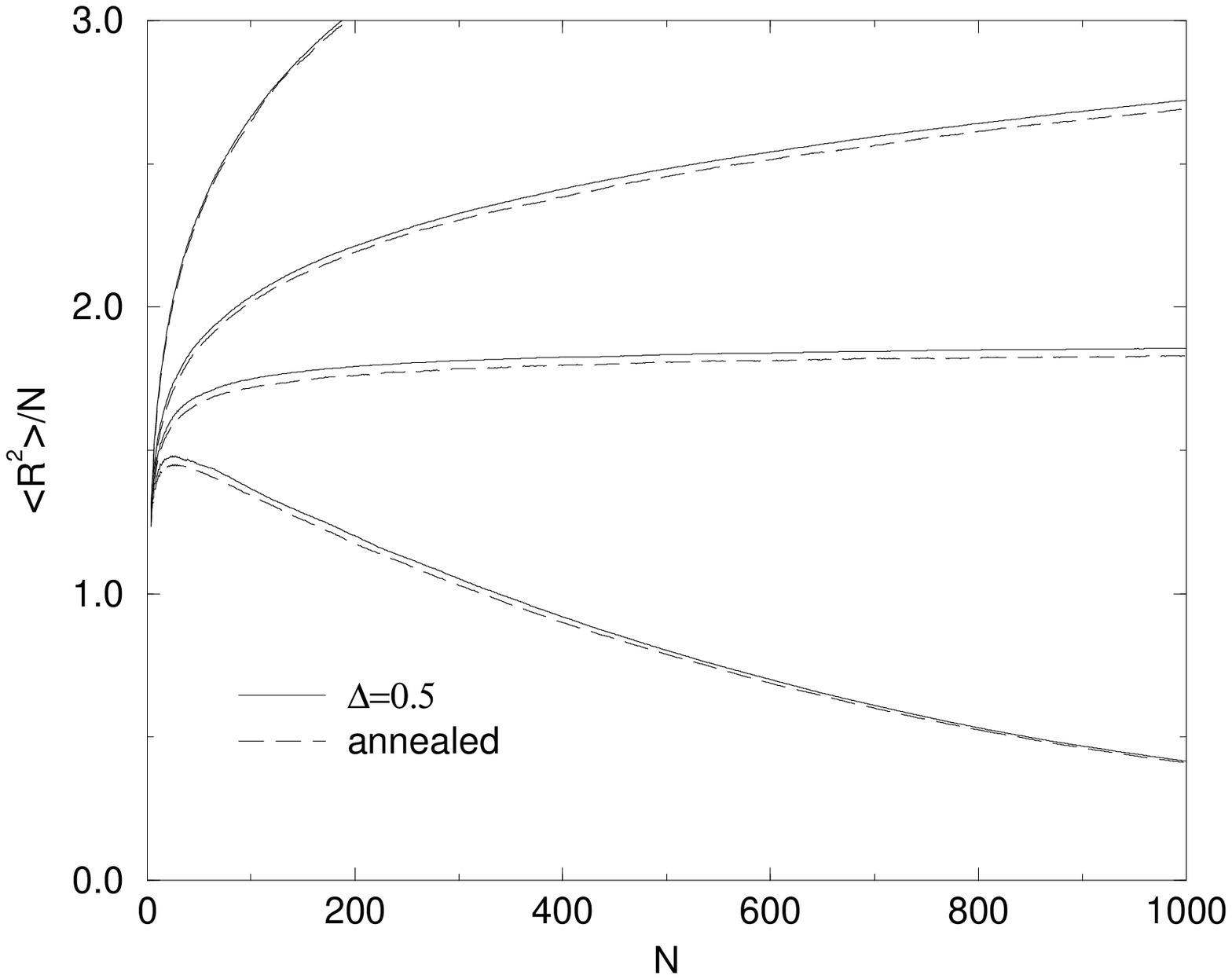,width=7cm}}

  \centerline{
    \psfig{file=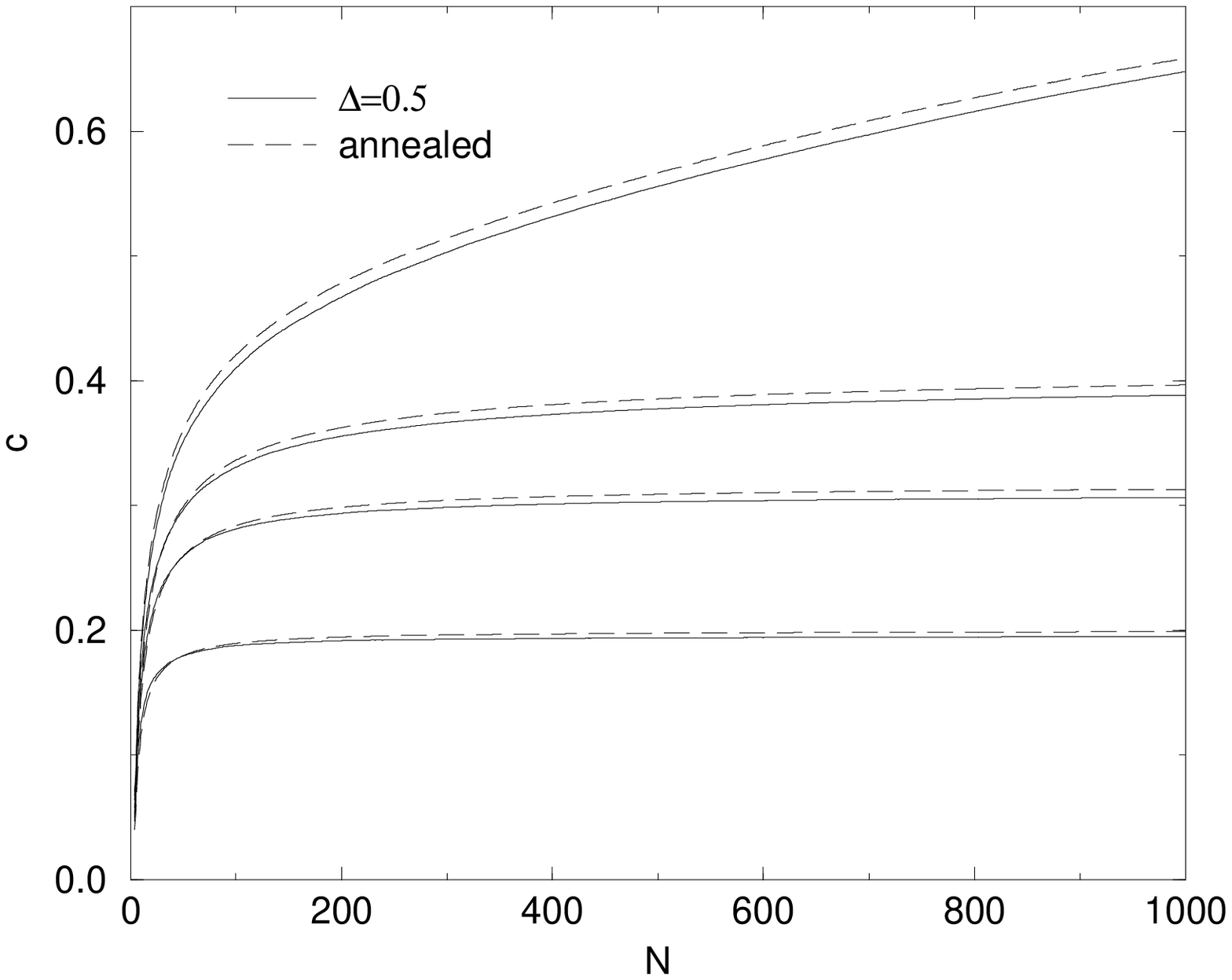,width=7cm}
    \psfig{file=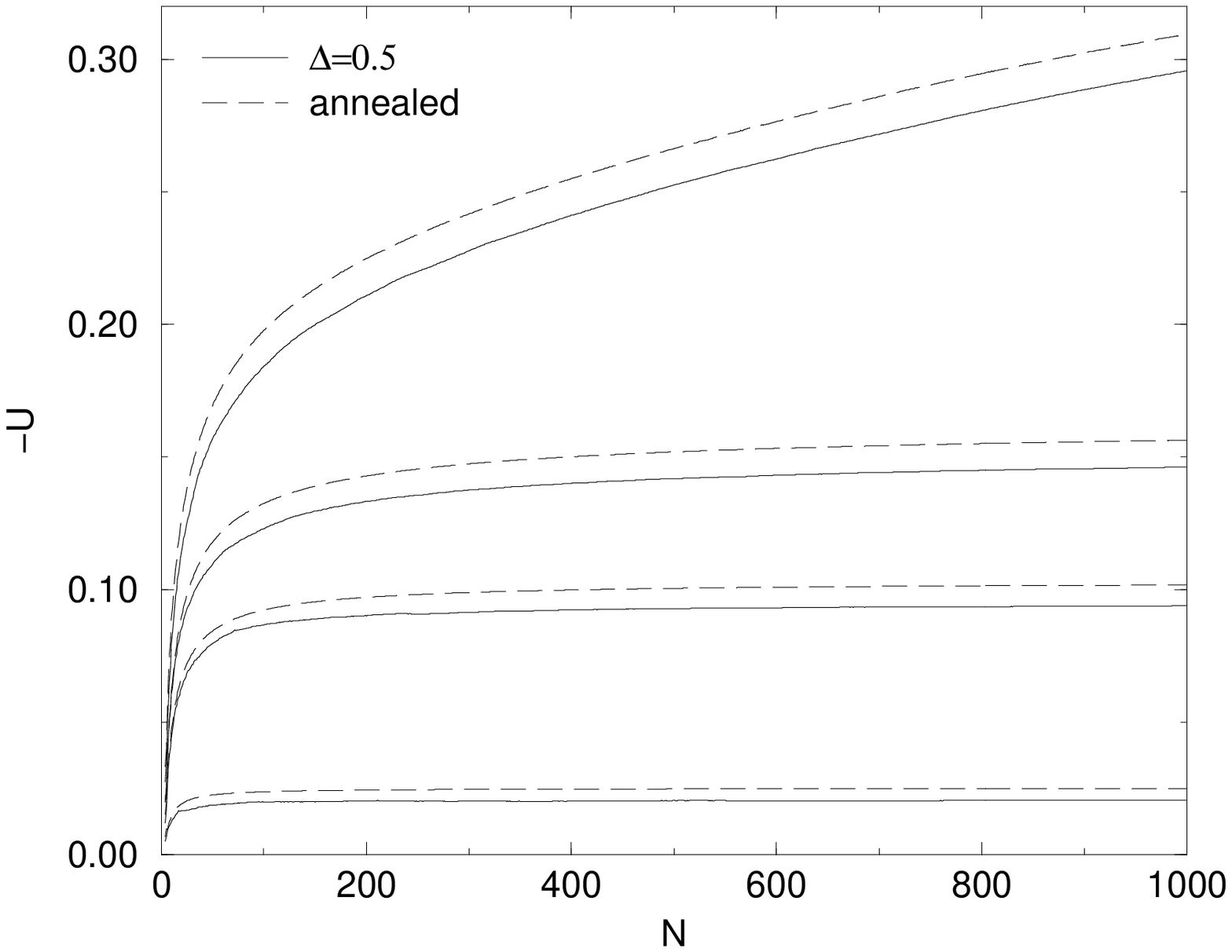,width=7cm}}
  \caption{Free energies $F_N$ per monomer (top left), end-to-end swelling 
    factors $\la R^2\ra/N$ (top right), density of contacts $c$ (bottom left), 
    and absolute values of energies per monomer $U_N$ (bottom right) 
    for four systems with $\D=0.5$, and with $\B=0, -0.20, -0.269$, and 
    $-0.345$. In the two top figures $|\B|$ increases from top to 
    bottom, in the bottom figures it increases from bottom to top. 
    Full lines are from Monte Carlo simulations, dashed lines are 
    predictions of the annealed approximation. In all panels, error 
    bars are much smaller than the thickness of the lines. The values 
    $\B=0$ and $-0.20$ are in the swollen phase, $-0.345$ is in the 
    collapsed phase, and $-0.269$ is on the theta line.}
  \label{fig:U_ann}
\end{figure}

The collapse line in the $(B,\D)$-plot obtained from simulating chains of 
length $N\leq 1000$ is shown in Fig.1. Here the solid line is the 
annealed approximation. We see that on this scale the annealed approximation 
seems perfect. But this is not quite true. 
Much more precise tests can be performed by comparing directly both sides of 
eqs.(\ref{F-annealed}) to (\ref{U-anneal}). Typical plots obtained in 
this way are shown in Figs.\ref{fig:U_ann}. For each 
of these plots we used at least 300 realizations of disorder, and we got 
at least $10^3$ independent configurations for each disorder realization. 
Similar plots were made also for several other values of $B$ and $\D$. 

In all 
these plots we see small but significant deviations. These deviations are 
present both in the collapsed (globular) and in the open (coil) phase. 
They depend only weekly on $N$. Therefore, even with our long chains and high 
statistics it is not clear whether they disappear for $N\to\infty$. 
Obviously, in order to proceed we need more refined theoretical predictions 
to compare with, and/or a more efficient way to do the disorder average.

Before we do this, we should point out that deviations from the annealed 
approximation were also found recently in a different model 
by Trovato {\it et al.} \cite{TMM}.

% -----------------------------------------------------------------------

\section{Replica Symmetric Approximation}

To go beyond the annealed approximation, we will use the replica trick
\be
   \ov{\ln Z_N}=\lim_{n\to 0}{\ov{Z_N^n}-1\over n}. \label{rep} 
\ee
Alternatively, we could try a Taylor expansion
\be
   \ov{\ln(Z_N/\ov{Z_N})} = - {1\over 2} (\ov{Z_N^2}/\ov{Z_N}^2 -1) +\ldots \;.
                 \label{taylor}
\ee
This expansion is most likely divergent. It is nevertheless useful 
since its first term gives already a good indication of the leading 
corrections. Also, it suggests the inequality
\be
   F_N(B,\D) \geq F_{N,\rm ann}(B,\D)
\ee 
which can easily be derived exactly from convexity of the logarithm.
The same inequality is expected to hold for $U_N$. This inequality
is equivalent to the existence of a negative correlation between the
average energy and the partition function.
%(but cannot be proven exactly) 

To use eq.(\ref{rep}), we first have to evaluate disorder averages of 
$Z_N^n$ for integer $n\geq 2$.
These are performed similarly to the average over $Z_N$, except that the 
Gaussian integrals give formally rise to interactions between replicas 
\cite{SG}, 
\be 
   \ov{Z^n}=\sum_{{\cal C}_1\cdots {\cal C}_n}\exp\l(
         -\B\sum_{\a=1}^n\l(\sum_{i<j}\s_{ij}^\a\r)+
         {\D^2\over 2}\sum_{\a\neq\b}\l(\sum_{i<j}\s_{ij}^\a\s_{ij}^\b\r)\r). 
                          \label{Zn}
\ee
Here the Greek indeces $\a$ and $\b$ refer to the different replicas, 
${\cal C}_\a$ is a configuration of replica $\a$, $\s_{ij}^\a$ is
its contact map and $\ti{B}$ is given by equation (\ref{ann2}).
The annealed approximation is equivalent to neglecting the two-replicas
term.

To proceed, we define the variables
\be
   c_\a={1\over N}\sum_{i<j}\s_{ij}^\a, \hspace{1cm}
    q_{\a\b}={1\over N\sqrt{c_\a c_\b}}\sum_{i<j}\s_{ij}^\a\s_{ij}^\b,
  \label{cq}
\ee
which are respectively the density of contacts for the contact map $\a$
and the overlap between two contact maps $\a$ and $\b$. The overlap is a
measure of similarity, and it is equal to one if and only if the two contact
maps coincide. Furthermore, we assume that, for large $N$, the
number of configuration $n$-tuples with $Nc_1,\ldots Nc_n$ contacts 
and mutual overlaps $\{q_{\a\b}\}$ grows as 
\be
\exp\l[N\l(\sum_{\a=1}^n f(c_\a)-\sum_{k=2}^{n}
\chi_k\l(\{c_\a\},\{q_{\a\b}\}\r)\r)\r].
%e^{N\l(\sum_{\a=1}^nf(c_\a)-\sum_{\a\neq\b}\chi_n(\{c^\a\},\{q_{\a\b}\})\r)}.
\ee
In other words, $\chi_k\l(\{c_\a\},\{q_{\a\b}\}\r)$ is the entropy loss
per monomer when we impose that the replica $\C_k$ with density of contacts
$c_\a$ has overlaps $q_{1,k}\cdots q_{k-1,k}$ with the $k-1$ previous
replicas. This quantity can be measured, for instance for $k=2$.

We can then write
\bea 
    & & \ov{Z^n}\approx \int d\{c_\a\}d\{q_{\a\b}\}\times \\  \nonumber
    & &\hspace{.5cm}\times \exp\l\{
          N\l[\sum_{\a=1}^n\l(f(c_\a)-\B c_\a\r)+{\D^2\over 2}\l(
       \sum_{\a\neq\b}\sqrt{c_\a c_\b}q_{\a\b}-
       \sum_{k=2}^n\chi_k\l(\{c^\a\},\{q_{\a\b}\}\r)\r)\r]\r\}\\ \nonumber
    & &\approx\hspace{.5cm}\exp\l[-NnF_n\l(\{c^\a\},\{q_{\a\b}\}\r)\r].
\label{exprep}\eea
Here, $F_n\l(\{c^\a\},\{q_{\a\b}\}\r)$ is the free energy per monomer 
in a system with $n$ replicas. To evaluate it, we approximate the integrals 
over $c_\a$ and $q_{\a\b}$ by their saddle points. We assume replica symmetry
which is expected to hold for low disorder: the saddle point is assumed to be 
given by $c_\a=c$ for all $\a$ and $q_{\a\b}=q$ for all pairs
$\a\neq\b$. 

Now, in order to obtain the correct free energy, we have to 
take the limit $n\to 0$. We obtain 
\be
   F(B,\D)=-f(c)+\B c+{1\over 2}\D^2 cq-\chi(c,q).  \label{F-saddle}
\ee
where $\chi(c,q)=-\lim_{n\to 0}\sum_{k=2}^n \chi_n(c,q)/n$ is the average
entropy gain per replica due to the condition that the overlap among all
replicas is equal to $q$. Note that this quantity is positive because, in
the limit of vanishing $n$, the number of terms in the sum is -1.
Finally, we have to compute the values of $c$ and $q$ at which $F$ is
evaluated by imposing two saddle point conditions:
 
\bea
  && {\p f(c)\over \p c} +{\p \chi(c,q)\over\p c}
	=\B+{1\over 2}\D^2q  \label{eq_c},
\\
   && {1\over 2}\D^2c = {\p \chi(c,q)\over\p q}.    \label{eq_q}
\eea

For $\D=0$ (homopolymer) Eq.(\ref{eq_q}) just means that the value of the
overlap is the most probable one for a given $c$, $q_0(c)$. Because of the
normalization, it must be $\chi(c,q_0)=0$,  thus the free energy
of the homopolymer is just a special case of Eq.(\ref{F-saddle}). Moreover,
since $\chi(c,q_0)=0$ is an absolute minimum, also the derivative
$\p \chi/\p c$ must vanish at that point, thus the saddle point equation
for $c$ valid for the homopolymer is recovered for $\D=0$. It also follows
from this argument that the second derivatives of $\chi(c,q)$ at the
point $(c, q_0(c))$ must be non-negative.

Notice that the free energy has to be
maximized as a function of $q$ because this variable refers to a space
with a negative number of dimensions in the limit $n\to 0$. We thus get
a first condition of thermodynamic stability ${\p^2 \chi/\p q^2}> 0$,
which, from the above consideration, is expected to be fulfilled for
$\D$ small enough.
The situation is more complicated for the variable $c$. It enters both
into the free energy of the replica interactions which has to be maximized
for $n\to 0$, and into the free energy of the homopolymer which has to be
minimized, at least for $\D=0$. We conjecture that the corresponding condition
of thermodynamic stability is that the Hessian determinant of the free
energy respect to the variables $c$ and $q$, $H(c,q)$, be non positive:

\be H(c,q)=\l({\p^2 f\over \p c^2}+ {\p^2\chi\over \p c^2}\r)
{\p^2\chi\over \p q^2}-
\l({\p^2\chi\over\p c\p q}-{1\over c}{\p\chi\over\p q}\r)^2 \leq 0.
\label{Denom}\ee

For $\D=0$ we have $H(c,q)=\l(\p^2 f/\p c^2\r)\l(\p^2 \chi/\p q^2\r)\leq 0$
as for homopolymers.
In fact, at that point the first derivatives of $\chi(c,q)$ vanish. The
Hessian determinant vanishes also, because $\chi(c,q)$ stays constant at
the value zero along the line $q=q_0(c)$. Thus both conditions of
thermodynamic stability are fulfilled at $\D$ small enough.

\begin{figure}[ht]
  \centerline{
    \psfig{file=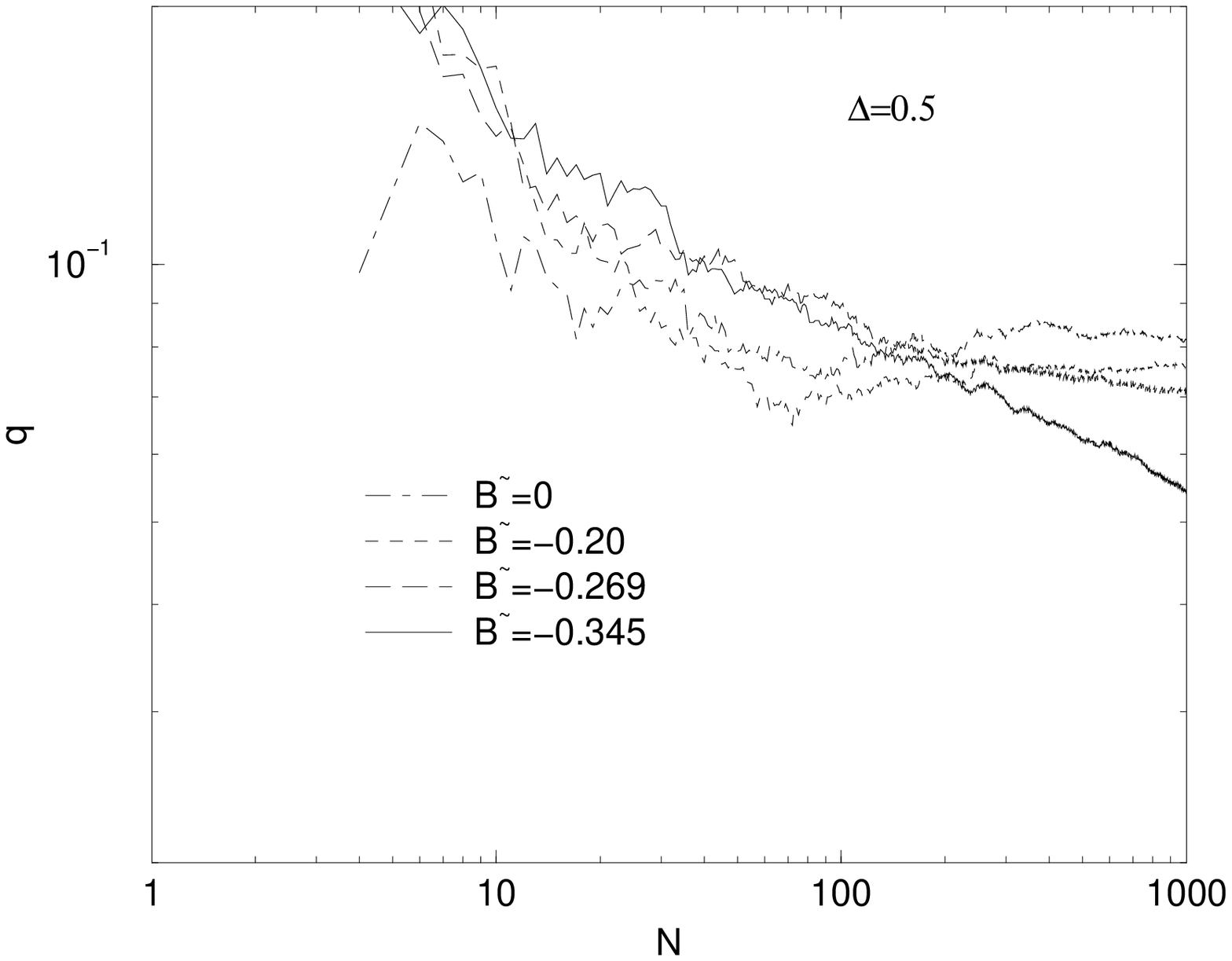,width=7cm}
    \psfig{file=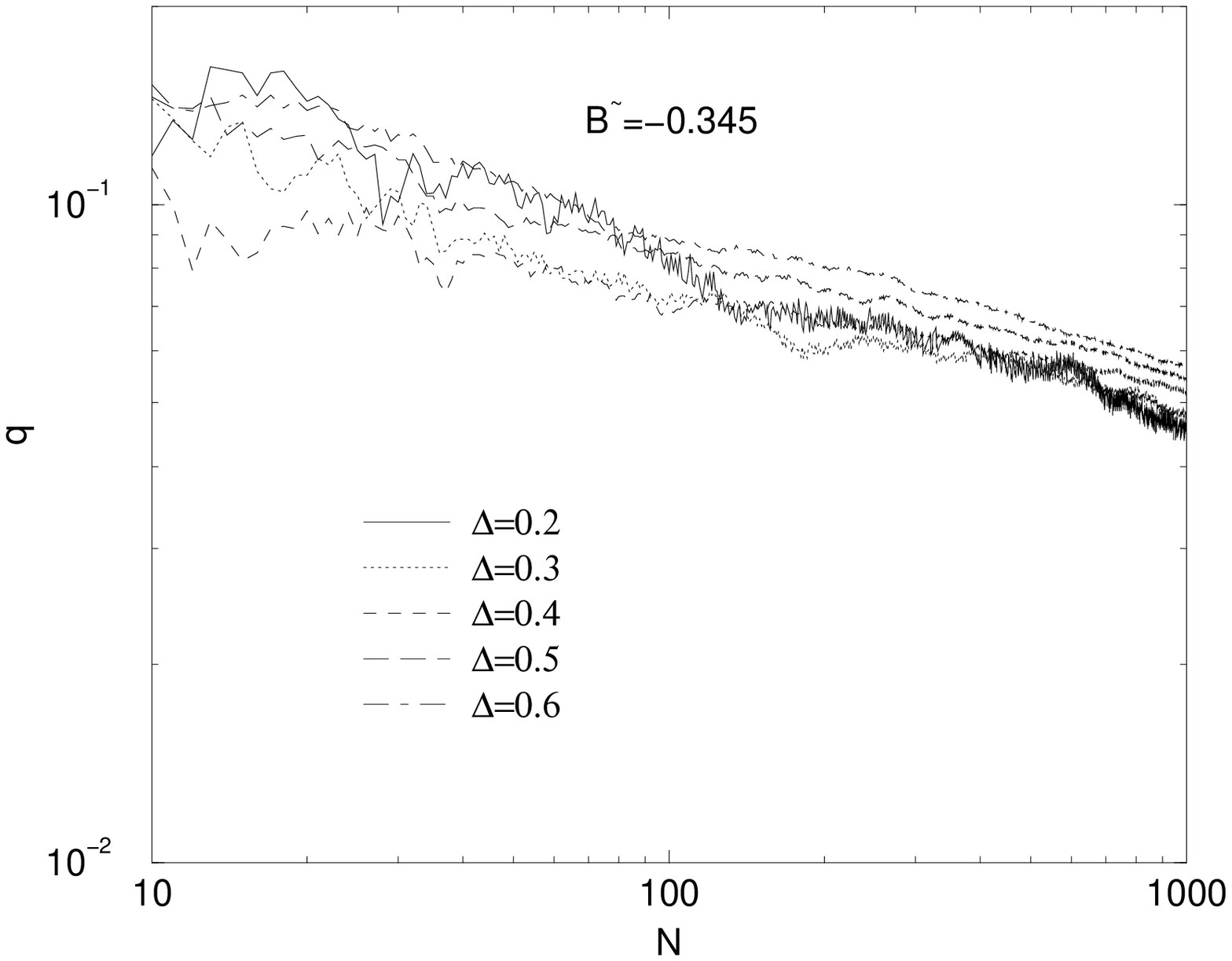,width=7cm}}
\caption{$q(B,\D)$ measured from the energy and the contact densities
    as a function of system size for $\D=0.5$ and four different 
    values of $=\B$ (panel a) and for $\B=-0.345$ and five different 
    values of $\D$ (panel b).}
\label{fig:q}
\end{figure}

The energy and entropy per monomer are obtained in the same way as in the
annealed  approximation. We find
\bea
   && U_N(B,\D) = \l(B-\D^2(1-q)\r)c, \label{Eq_Ene}\\
   && S_N(B,\D)= f(c)+\chi(c,q)-{\D^2\over 2}c(1-q). \label{Eq-Ent} 
\eea
Although this is obtained from the saddle point method which is 
exact only for $N\to\infty$, we can use Eq.(\ref{Eq_Ene}) to obtain effective
values of $q$ also for finite $N$. Results are shown in Fig.\ref{fig:q}a.
It appears that $q$ decreases with system size, but its asymptotic value
seems to be finite in the random coil phase. This was confirmed by similar
measurements at different values of $\Delta$. Thus the annealed approximation
does {\it not} hold in the random coil phase.
The situation is more difficult for collapsed chains. In this case it can
not be excluded from Fig.\ref{fig:q}b that the overlap asymptotically
vanishes, and thus the annealed approximation becomes exact in the
thermodynamic limit. However, simulations of systems of even larger size
that will be presented in the next section show that this is not
the case and that the corrections to the annealed approximation remain
finite in the thermodynamic limit in both the coil and the collapsed phases,
although in the latter phase they are rather small. As seen from 
Fig.\ref{fig:q}b, $q$ does not depend
very much on $\D$ in the collapsed phase and for large systems.

\begin{figure}[ht]
  \centerline{
    \psfig{file=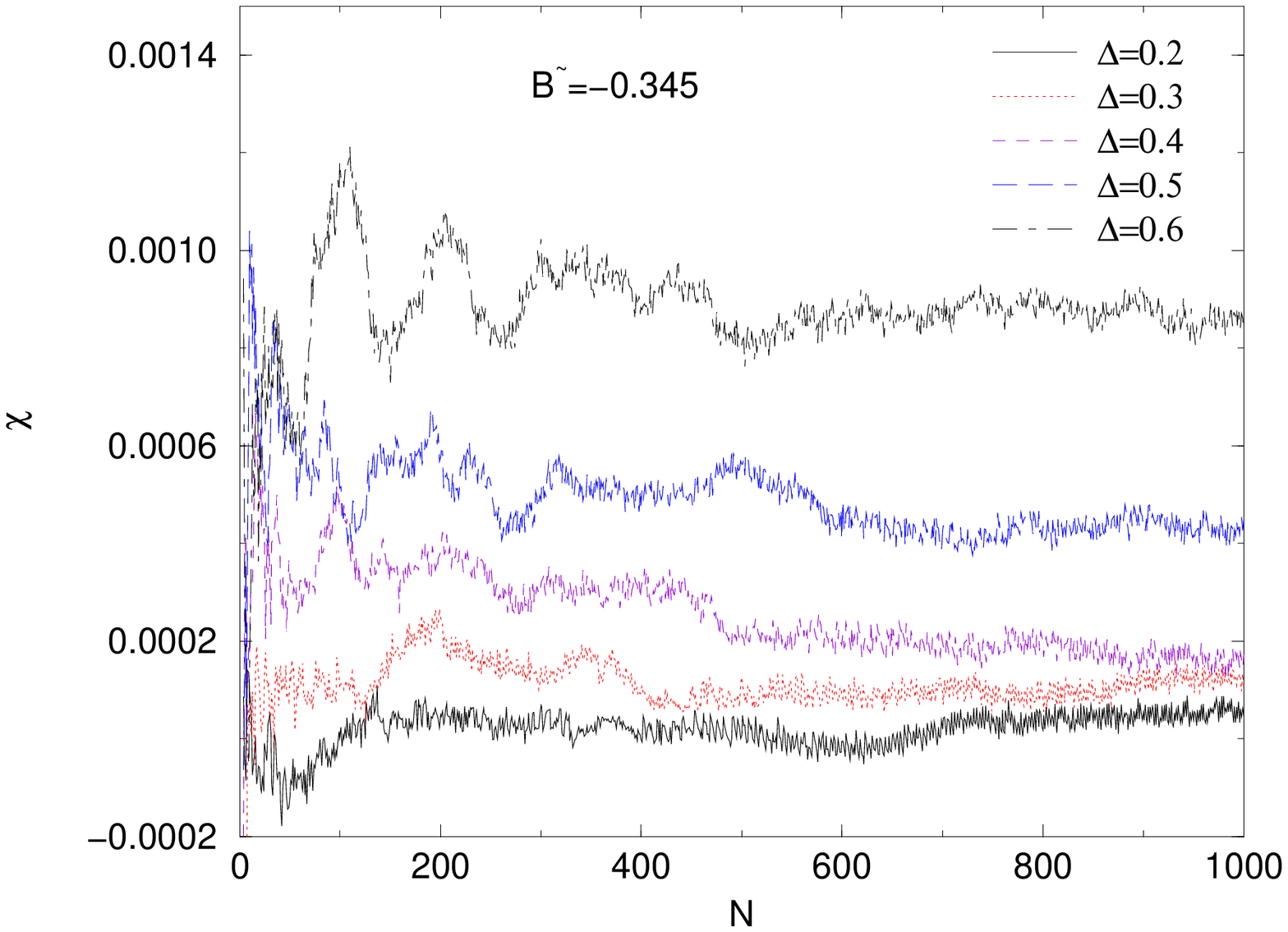,width=7cm}
    \psfig{file=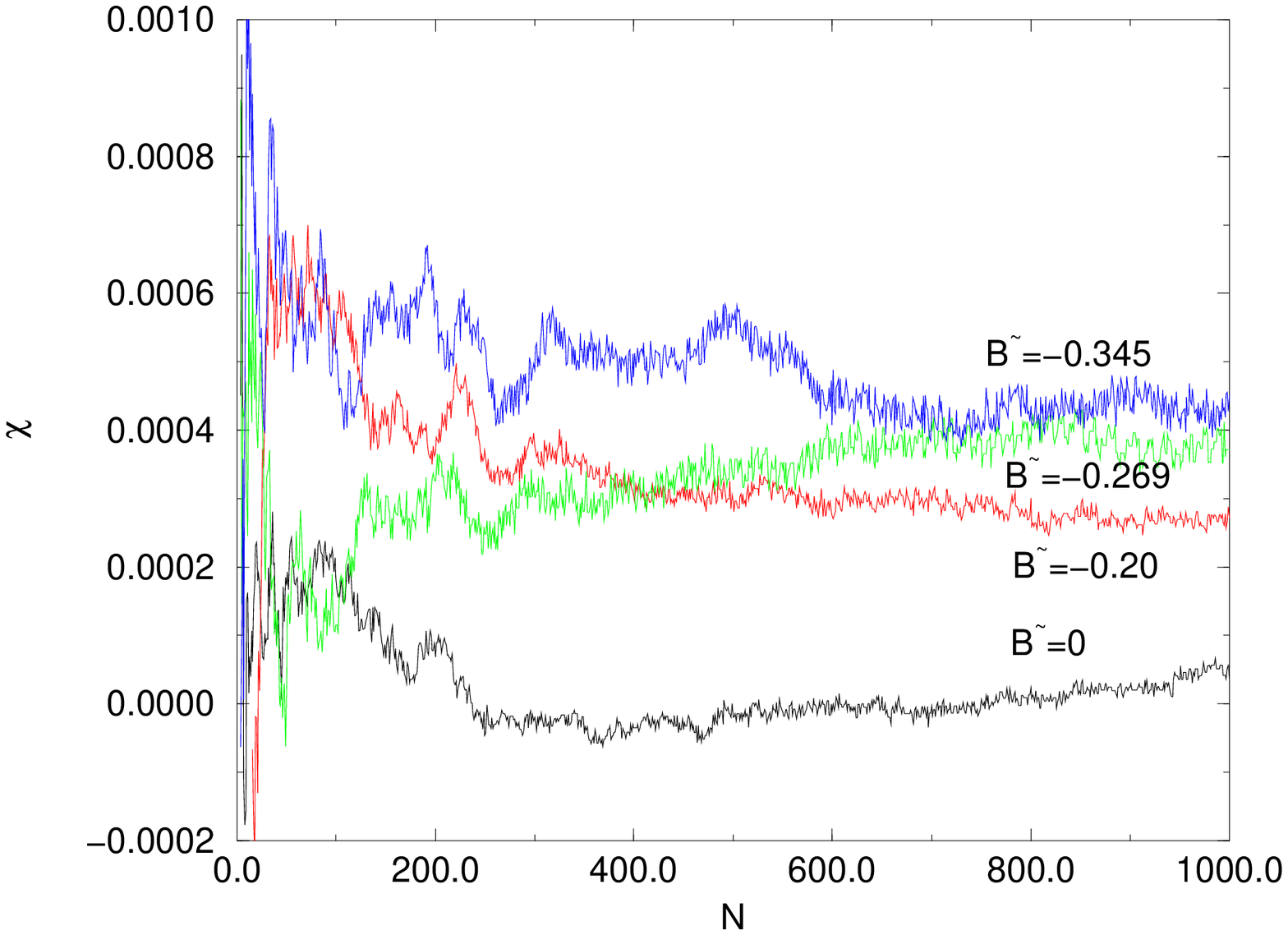,width=7cm}}
\caption{$[\chi(B,\D)+\chi(\B,0)]/2$ measured from Eq.(\ref{eq-chi})
    as a function of system size for the same values of $\B$ and 
    $\D$ as in Fig.\ref{fig:q}.}
\label{fig:chi}
\end{figure}

To obtain numerical estimates of $\chi$, we subtract from 
Eq.(\ref{F-saddle}) the analogous equation for homopolymers, 
and obtain 
\be
   F(B,\D)-F(\B,0) = (c-\c)\B - [f(c)-f(\c)]+{1\over 2}\D^2cq -\chi(c,q).
\ee
Expanding the difference $f(c)-f(\c)$ in the difference $c-\c$ 
and using $f'(\c)=\B$, we get 
\be
   F(B,\D)-F(\B,0) = -{1\over 2}(c-\c)^2f''(\c)+{1\over 2}\D^2cq 
      -\chi(c,q) +O\l((c-\c)^2\r).      \label{dF}
\ee
This can be either evaluated directly, neglecting the last term. 
Alternatively, we can eliminate the term involving $f''(\c)$ by 
subtracting from Eq.(\ref{eq_c}) the analogous equation for homopolymers, 
which gives 
\be
   (c-\c)f''(\c) = {1\over 2}\D^2q-{\p \chi(c,q)\over\p c} 
        +O\l((c-\c)^2\r) .                           \label{df}
\ee
Combining this with Eq.(\ref{dF}) and neglecting terms $O\l((c-\c)^2\r)$, 
we obtain finally
\be
   F(B,\D)-F(\B,0) = {c+\c\over 4}\D^2q -{1\over 2}[\chi(c,q)+\chi(\c,q)]
       +O\l((c-\c)^2\r) .
       \label{eq-chi}
\ee
In Fig.\ref{fig:chi} we show numerical values for $[\chi(c,q)+\chi(\c,q)]/2$ 
obtained in this way. We see that $\chi$ is very small but definitely 
not zero. Again we see that corrections
to the annealed approximations are larger in the swollen 
phase than in the collapsed phase. Indeed, this time it seems that 
the deviations have a finite limit for $N\to\infty$ in both phases.
This conclusion is supported by measurements at other values of $\D$ (not
shown here).  Basically the same conclusions are also 
drawn from Eq.(\ref{dF}), showing that $\chi(c,q)$ depends weakly on $c$ 
and that Eq.(\ref{df}) is very well satisfied.

\section{Overlap of Homopolymers}

In this section we will discuss the overlap of contact matrices for 
homopolymer chains and their relation to corrections to the annealed 
approximation in the weak disorder limit. Consider the derivative of the 
free energy of a random heteropolymer with respect to $\D^2$, at 
$\D=0$. Using Eqs.(\ref{taylor},\ref{Zn}) and the results of sec.2, we obtain
\be 
   \left[{\p F_N(B,\D)\over \p \D^2}\right]_{\D=0} = 
          N^{-1} \left[2{\p \ln \ov{Z_N} \over \p \D^2}
                     - {1\over 2}{\p \ov{Z_N^2}\over \p \D^2}\right]_{\D=0}
         = -{1\over 2}{\p F_N(B,0)\over \p B} + {1\over 2N}
            \sum_{\C_1,\C_2}\sum_{i<j} \sigma^1_{ij}\sigma^2_{ij} \;.
\ee
Thus we have 
\be
    \left[{\p\over \p \D^2}\left( F_N-F_{N,\rm{ann}}\right)\right]_{\D=0}
            = {1\over 2} \langle cq\rangle .
       \label{pair}
\ee
This could have been obtained of course also within the replica symmetric 
approach, but the above derivation shows that it is indeed a rigorous result 
involving neither approximations nor unjustified assumptions.

Notice that this cannot be generalized to $\D\neq 0$, but in principle 
straightforward generalizations could be used to compute all higher derivatives 
\be
    \left[{\p^k\over \p \D^{2k}} \left(F_N-F_{N,\rm{ann}}\right)\right]_{\D=0},
       k = 1,2,3,\ldots \;.
\ee

Numerically, the rhs. of Eq.(\ref{pair}) can be estimated by simulating pairs 
of chains simultaneously. For this we used a variant of the PERM algorithm 
where we add monomers alternatively to the first and to the second chain
\cite{CCG}. In this way we guarantee that both chains have exactly the same 
length (after having added an even number of monomers), and it is 
straightforward to estimate their overlap.

\begin{figure}[ht]
  \centerline{
    \psfig{file=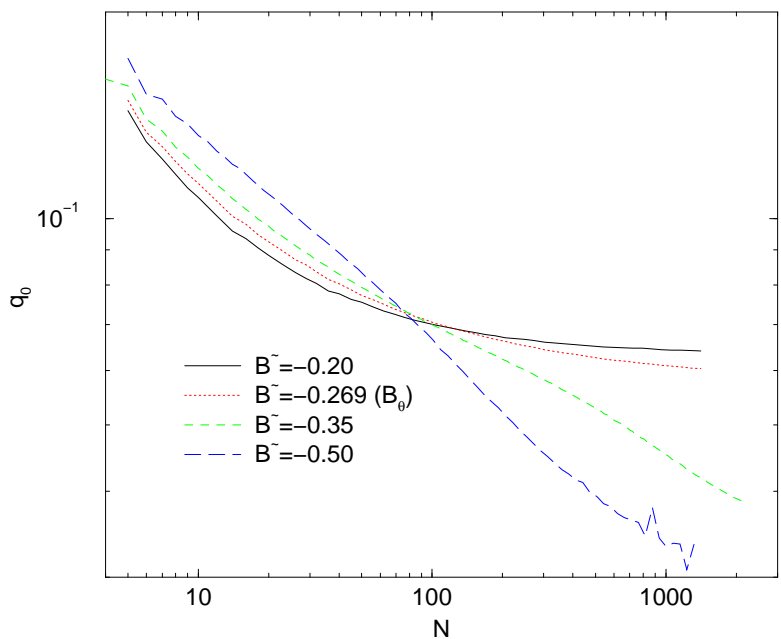,width=12cm}}
\caption{Average overlap $q_0 = \langle cq\rangle/\langle c\rangle$ 
between homopolymer chains of length $N$ for
different values of the monomer-monomer attraction $B$.}
\label{fig:q0}
\end{figure}

Results from such simulations with chains of length up to 1400 are shown in 
Fig.\ref{fig:q0}. These data agree nicely with extrapolations of 
the overlaps for $\D>0$ shown in the last section. They have much smaller 
statistical errors, since we do not have to average over any disorder 
explicitly. This makes the present method much faster and allows us to 
study larger systems.

\begin{figure}[ht]
  \centerline{
    \psfig{file=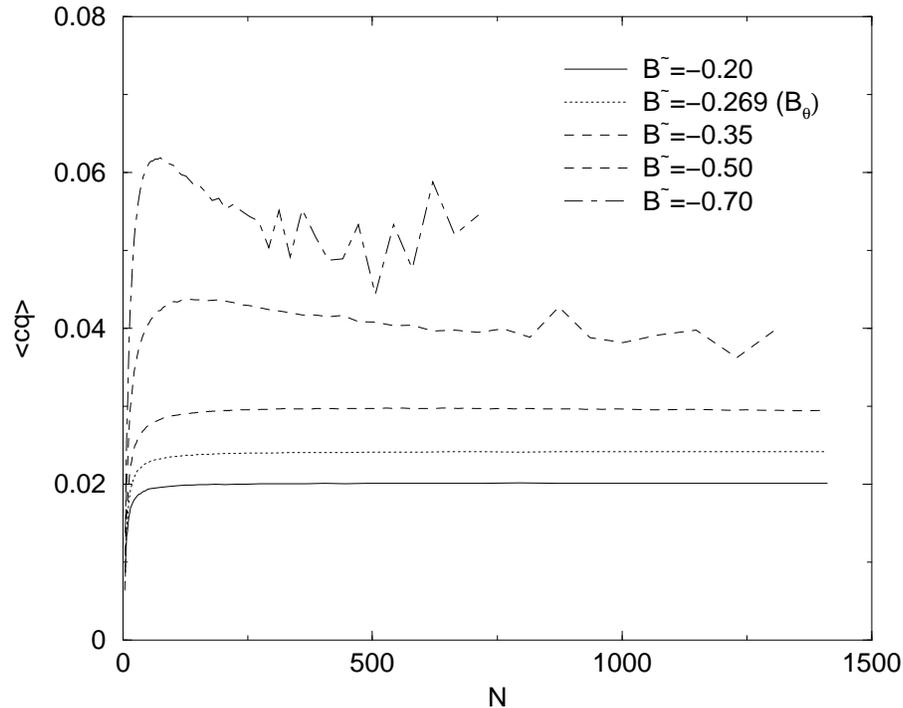,width=12cm}}
\caption{Average fraction of common contacts $\langle cq\rangle$ 
between homopolymer chains of length $N$ for
different values of the monomer-monomer attraction $B$.}
\label{fig:cq}
\end{figure}

The curve for $\B=-0.2 > B_\th$ in Fig.\ref{fig:q0} shows clearly that 
the annealed approximation does not become exact for $N\to\infty$ in the 
open coil phase. The same is true (although a bit less clear) exactly at 
the $\Theta$ point, as indicated by the curve for $\B=-0.269$. For the 
collapsed phase, the evidence is not so clear. The curves for $\B=-0.35$ 
and for $\B=-0.5$ both are much lower for large $N$ and continue to 
decrease. Superficially, one might therefore conclude that the overlap 
disappears for $N\to\infty$. But both these curves show a distinct upward 
curvature for the largest values of $N$, indicating that the decrease 
will level off and $q_0$ tends  to a finite constant for $N\to\infty$. To
sustain this view, we show in Fig.\ref{fig:cq} the plots of $\langle cq\rangle$
as a function of chain length $N$. In this case it is evident that the curves
are going to a non-zero value. Since the fraction of contacts is limited (it
holds $c\leq 2$ on the cubic lattice), also $q_0 = \langle cq\rangle/
\langle c\rangle$ should go to a non-zero value, and the decrease observed
in Fig.\ref{fig:q0} is just a consequence of the fact that the average
fraction of contacts is increasing, approaching its stationary value.

This conclusion is not, after all, very surprising. It can be backed by 
an argument which could presumably, with some effort, be made rigorous.
The density of contacts in a self-avoiding walk (SAW, corresponding 
to $B=0$) is dominated by short range contacts, i.e. by contacts $(i,j)$
with small $|i-j|$. The contribution of such a loop with fixed $i$ and 
$j$ depends weakly on the configuration of the chain far away from 
this monomer pair. Thus, if such a loop is present in one replica, it 
has a high chance to be present also in the other replica, even if the 
global structures of both replicas are entirely different.

Notice that this argument depends crucially on the fact that it is the 
overlap between {\it contact matrices} which is of relevance, not the 
geometric overlap between configurations. It seems that the authors 
of \cite{MSVO} have missed this in an otherwise very similar argument
which has led them to the opposite conclusion that the annealed 
approximation does become exact.

In collapsed chains there are relatively more long range contacts, which 
explains why the annealed approximation is better -- but still not 
exact -- in that regime. For random heteropolymers with strong disorder,
and even more so for proteins, this argument suggests an increased 
overlap because of secondary structure. 
For instance, an alpha helix produces an array of contacts $(i,i+4)$,
$(i+1,i+5), \cdots$, where $i$ labels the position of the amino acid along
the protein chain. This enhances the overlap. Moreover, there is a 
finite probability that such contacts appear simultaneously even in the 
structures of two unrelated proteins. Thus the average overlap even 
in a large set of unrelated protein structure appears to attain a
finite limit when the length of the chains increases \cite{UMW}.

\section{Thermodynamics in the replica symmetric approximation}

The two saddle point equations for $c$ and $q$ can not be explicitly
solved, because we lack an explicit expression for the functions $f(c)$
and $\chi(c,q)$. Nevertheless, their qualitative behavior can be studied in
more detail. This is done in the present section.

Deriving both equations (\ref{eq_c},\ref{eq_q}) with respect to the
thermodynamic parameters $\B$ and $\D$ we can compute the
derivatives of $c$ and $q$ as

\bea 
\l( {\p c\over \p\B} \r)_{\D} = { {\p^2\chi\over\p q^2} \over H(c,q) }
\leq 0, \:
& &
\l( {\p c\over \p\D} \r)_{\B} = {-\D\over H(c,q)} {\p\over \p q}
\l( c {\p \chi\over \p c} -q {\p\chi\over\p q} \r). \label{der}
\\
\l({\p q\over \p\B}\r)_{\D} =
-{{\p^2\chi\over\p c\p q}-{1\over c}{\p\chi\over\p q} \over H(c,q)},
& &
\l({\p q\over \p\D}\r)_{\B} = {\D\over H(c,q)}
\l[ c \l({\p^2 f\over \p c^2}+ {\p^2\chi\over \p c^2}\r)-
q \l({\p^2\chi\over\p c\p q}-{1\over c}{\p\chi\over\p q} \r)\r].
\nonumber\eea
where $H(c,q)$ is given by Eq.(\ref{Denom}). From these, the specific heat
can be computed as

\bea
C_v & = & {\p U/\p T} \nonumber\\
    & = &
\l(B-\D^2(1-q)\r)\l[(B-\D^2){\p c\over\p\B}+\D{\p c\over\p\D}\r]
 -\D^2c\l[(1-q)+(B-\D^2){\p q\over\p\B}+\D{\p q\over\p\D}\r]
       \nonumber\\
& = & 
-{1\over H(c,q)}\l[{\p^2F\over \p c^2}(\D^2c)^2-
2{\p^2F\over \p c\p q}(\D^2c)(B-\D^2(1-q))+
{\p^2F\over \p q^2}(B-\D^2(1-q))^2 \r] \nonumber\\
&  & +\D^2c(1-q),
\eea
where $F(c,q)$ is the free energy evaluated at the saddle point.
The three terms in the square brackets are the quadratic form whose
determinant is expressed by $H(c,q)$. Since $\p^2 F/\p q^2$ is positive,
they would give a
negative contribution to the specific heat if $H(c,q)$ were positive. 
Thus it is justified to require that $H(c,q)$ is negative in order to
get thermodynamic stability.

The behavior of the thermodynamic derivaties (\ref{der}) can be
partly studied using the fact that the function $\chi(c,q)$ attains
its absolute minimum value $\chi(c,q)=0$ along the line $q=q_0(c)$. Thus,
assuming that it is an analytic function of $q$ for $q>0$, it can be
expressed in the form
\be
\chi(c,q)=\sum_{k=2}^\infty {a_k(c)\over k!}\l(q-q_0(c)\r)^k
\equiv\sum_{k=2}^\infty {A_k(c)\over k!}\l(Q-Q_0(c)\r)^k\:,
\label{chi_approx}
\ee
with $a_2(c)>0$. The typical overlap $q_0(c)$ is a small quantity and
it is a decreasing function of $c$, or $q'_0(c)<0$ (the prime indicates
derivative with respect to $c$). The coefficients $a_k(c)$ are expected
to be quantities of order $\l(q_0(c)\r)^{-k+1}$, as it will be argued later.
We also introduce the notation $Q=cq$, $Q_0(c)=cq_0(c)$,
$A_k(c)=a_k(c)c^{-k}$. 
We can now develop the saddle point equations for $c$ close to the
solution $\c$ of the annealed approximation given by $f'(\c)=\B$:

\bea
& & \l(
{\p^2 f\over c^2}+{\D^2\over 2}{\p^2 Q_0\over\p c^2} \r)_{c=\c}
\l(c-\c\r)+
\l({\D^2\over 2}{\p Q_0\over\p c} \r)_{c=\c}+
\l(\sum_{k=2}^\infty{A'_k(c)\over k!}
\l(Q-Q_0(c)\r)^k\r)_{c=\c}=0 \nonumber \\
& &\sum_{k=1}^\infty {A_{k+1}(c)\over k!}\l(Q-Q_0(c)\r)^k={1\over 2}\D^2\:.
\eea
From these expressions one sees that both $Q-Q_0(c)$ and $c-\c$ are quantities
of order $Q_0(c)$, thus corrections to the annealed approximation are small but
finite for finite $\D$. Moreover, $\d q=(q-q_0(c))$
is positive for small $\D$. From Eq.(\ref{chi_approx}) we find,
to zero-th order in $\d q$:

\bea  
\l({\p c\over \p\B}\r)_{\D} \approx {a_2(c)\over H(c,q)} \leq 0\:\:
& &
\l({\p c\over \p\D}\r)_{\B} \approx {\D a_2(c)\over H(c,q)}
{\p Q_0\over \p c}
\\
\l({\p q\over \p\B}\r)_{\D} \approx {a_2(c)\over H(c,q)}
{\p q_0\over \p c}\geq 0
& &
\l({\p q\over \p\D}\r)_{\B} \approx  {\D \over H(c,q)}
\l[c {\p^2 f\over \p c^2}+a_2(c){\p q_0\over \p c}{\p Q_0\over \p c}\r]
\label{chi(cq)}
\eea
$H(c,q)$ must be computed at the first order in $\d q$, because the
zero-th order term vanishes at the theta point $c=c_\th$ at which
$\p^2 f/\p c^2$ vanishes. The result reads:

\be H(c,q)\approx {\p^2 f\over \p c^2}{\p^2 \chi\over \p q^2}-
\l(Q-Q_0(c)\r)c(A_2(c))^2{\p^2 Q_0\over \p c^2}.
\label{HH}\ee

To proceed further, we first consider the simple case where the
entropy $\chi(c,q)$ depends only on the product $Q=cq$ representing the
number of conditions that we have to impose in order to fix an overlap $q$:
$\chi(c,q)=\ti\chi(cq)$.
This form was assumed in the work of Shakhnovich and Gutin \cite{SG}.
As it is easy to see, under the above hypothesis $c=c(\B)$ coincides with
the value predicted by the annealed approximation and it is independent
of $\D$ at fixed $\B$. Thus, in this case the annealed
approximation would yield the correct value for the fraction of contacts.
However, our numerical results contradict this prediction.
The overlap $q$ is in this case an increasing function of $\D$, and its
derivative with respect to $\B$ is proportional
to $\l({\p^2 f/\p c^2}\r)^{-1}$, which is expected to
diverge at the theta point $c=c_\th$. This fact can explain why the values of
$q$ are much smaller in the collapsed phase than in the coil phase.

\begin{figure}[ht]
  \centerline{
    \psfig{file=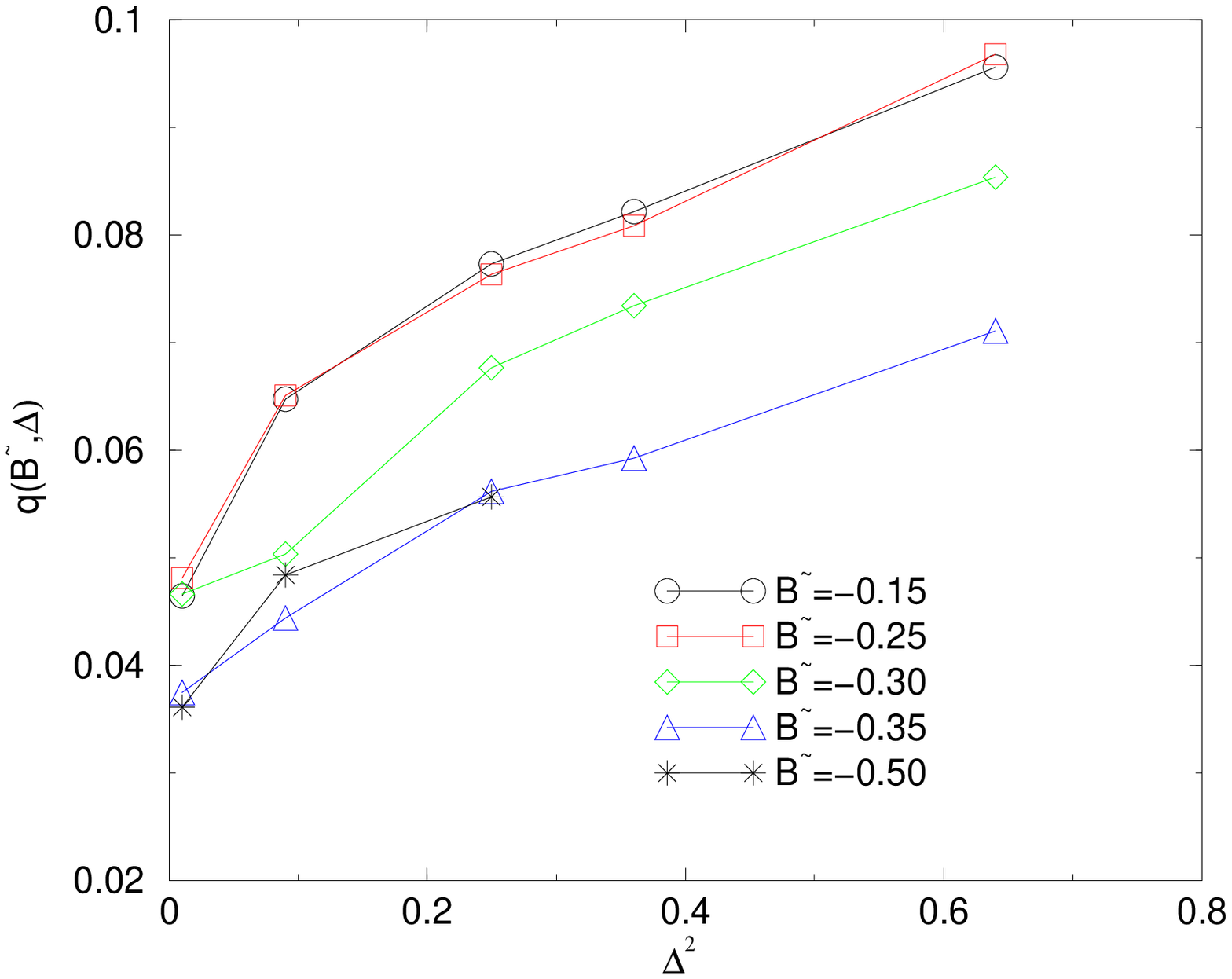,width=7cm}
    \psfig{file=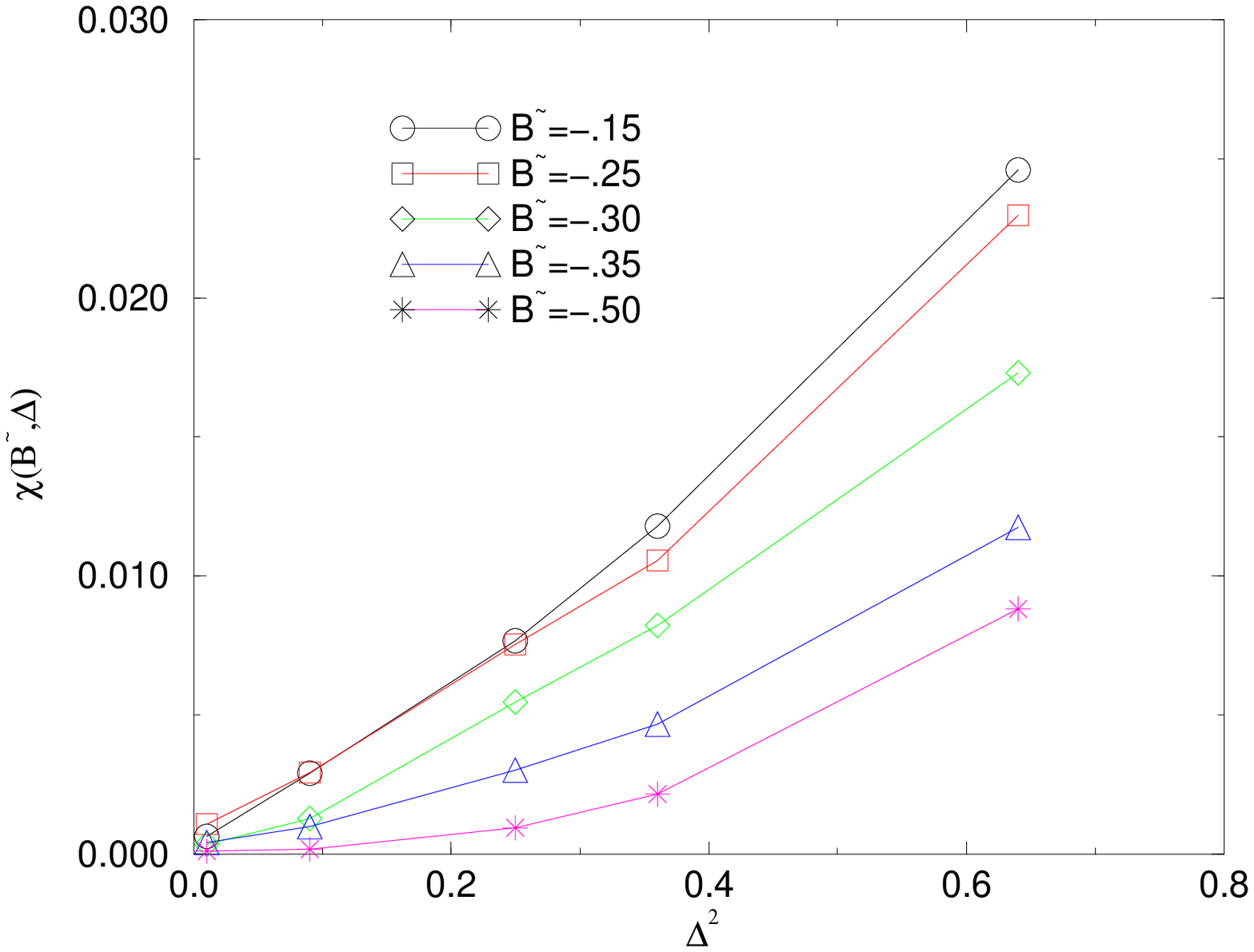,width=7cm}}
\caption{Left: overlap $q$ as a function of $\D^2$ for different values
of $\B$. Right: entropy $[\chi(B,\D)+\chi(\B,0)]/2$ measured from Eq
(\ref{eq-chi}) as a function of $\D^2$ for different values of $\B$.}
\label{fig:delta}
\end{figure}

\begin{figure}[ht]
  \centerline{
    \psfig{file=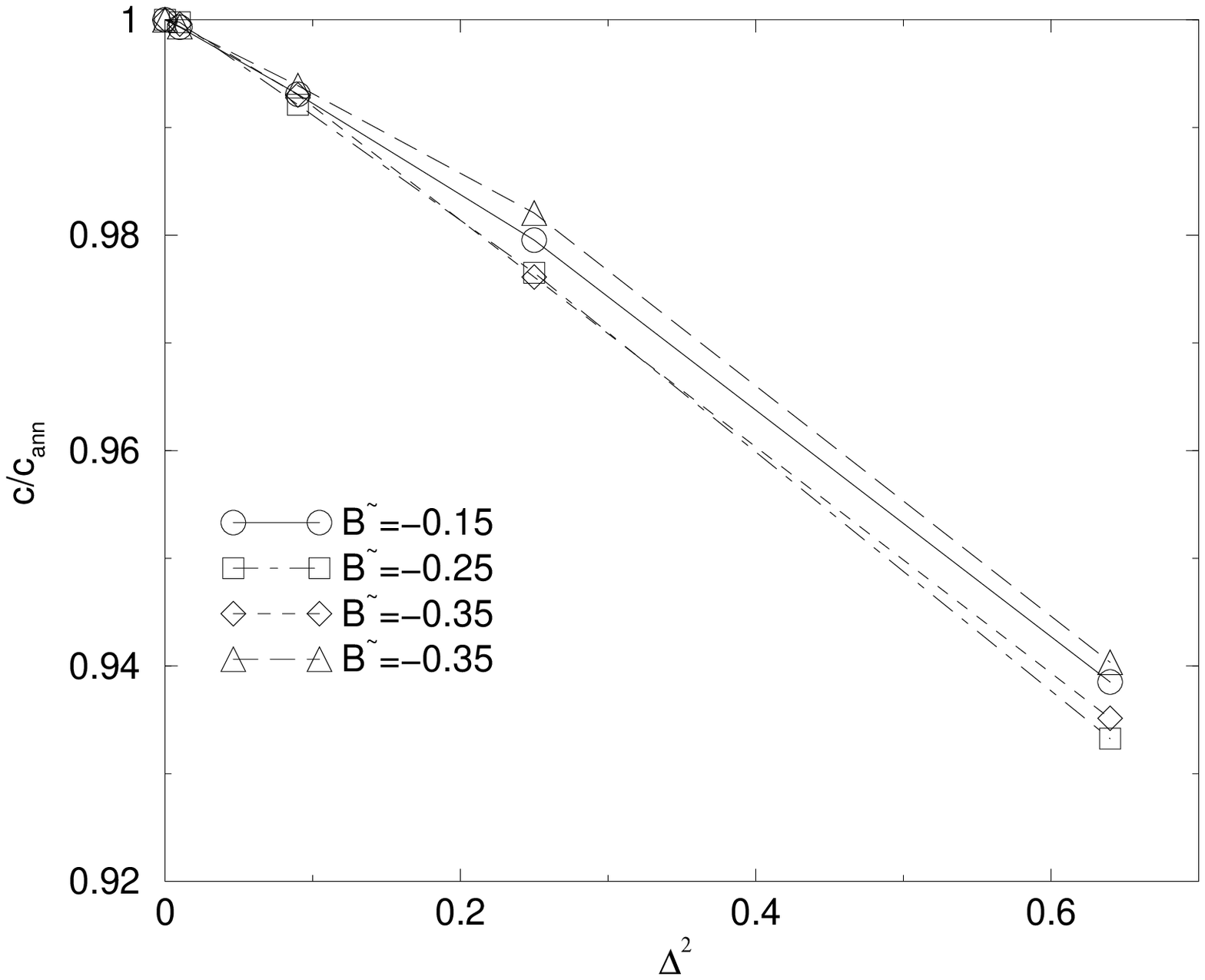,width=12cm}}
\caption{Corrections to the density of contacts predicted by the
annealed approximation as a function of $\D$, for different
values of $\B$ and $N=800$. The deviations are maximal close to the theta
point $\B\approx -0.27$.}
\label{fig:c_delta}
\end{figure}

A better theory for the function $\chi(c,q)$ was developed by
Plotkin, Wang and Wolynes \cite{PWW}. They computed the entropy loss for two
replicas with density of contacts $c$ being at overlap $q$, $\chi_2(c,q)$,
in the mean field approximation and in the collapsed phase. Although this
can be different from $\chi(c,q)$,
it is a good point for understanding its qualitative behavior. Unfortunately,
we can not use the formula obtained in \cite{PWW} because of several reasons:
first, the result is valid for finite $N$ and becomes pathologic as
$N\to\infty$; second, it is not given in the form of a differentiable
function; and third, it is assumed that the most probable overlap $q_0(c)$
is zero in the thermodynamic limit, while our calculations show that this is
not the case. We shall however
use the fact that $\chi(c,q)$ is the sum of three contributions:
the entropy loss due to imposing that $Ncq$ contacts have to coincide, the
loss due the fact that $Nc(1-q)$ contacts have to be different, and a
combinatoric factor counting the number of different choices of $Ncq$
contacts among $Nc$. The last term can be approximated by
$\chi_{\rm mix}(c,q)=c\l[q\ln q+(1-q)\ln (1-q)\r]$, even if this is
an overestimate, since not all of the combinations of different common
contacts can be realized. Putting everything together we have:

\be \chi(c,q)=\ti\chi(c,cq)+c\l[q\ln q+(1-q)\ln (1-q)\r]\: \label{Plot}.
\ee
In the computation by Plotkin {\it et al.} the mixed second derivative of
$\ti\chi(c,cq)$ with
respect to $Q=cq$ and $c$ vanishes. This simplifies considerably formulas,
and it will be assumed to hold for the rest of the paper. We shall thus
introduce the notation $\ti\chi^\pr(Q)$ to denote the derivative of
$\ti\chi(c,Q)$ with respect to $Q=cq$ at fixed $c$. Comparing Eq.(\ref{Plot})
to Eq.(\ref{chi_approx}), we see that $\ti\chi^{\pr\pr}(Q)$ must be
positive and that $\p\ti\chi/\p c=-(1-2q_0)\log(1-q_0)$.
We also see that
$a_2(c)=c/(q_0(1-q_0))+\ti\chi^{\pr\pr}(Q)$
is likely to be a quantity of order
$q_0^{-1}$, as it has been assumed above. For the higher order coefficients
one finds $a_k=O\l(q_0^{-k+1}\r)$, as anticipated. We have now to compute
the derivatives of $Q_0(c)=cq_0(c)$:

\be
{\p Q_0\over\p c}=-{\p^2\chi/\p c\p Q\over\p^2\chi/\p Q^2}=
%{c\over a_2(c)\l(1-q_0(c)\r)}
{q_0\over 1+cq_0\ti\chi^{\pr\pr}(cq_0)\l(1-q_0\r)}
\geq 0\: , \label{Q}
\ee
in qualitative agreement with Fig.\ref{fig:cq}.
Inserting the above result in the formulas (\ref{chi(cq)}) we see that the
density of contacts decreases with $\D$ at fixed $\B$. This behavior is
confirmed by our numerical results (see Fig.\ref{fig:c_delta}), which also
show that the decrease is maximal for $\B\approx B_\th=-0.27$, as expected
from the fact that $H(c, q_0(c))$ vanishes at $c=c_\th$. 
The overlap $q$ increases with $\B$ at fixed $\D$,as expected from
Eq.(\ref{chi(cq)}) (see Fig.\ref{fig:delta}), and increases
with $\D$ at fixed $\B$, as expected from Eq.(\ref{Q}) (see
Fig.\ref{fig:delta} again). It can thus be understood
why the overlap decreases with system size: as the number $N$ of monomers
increases the importance of surface effects is reduced (as $N^{-1/3}$) and
$c(N)$ increases, thus decreasing the value of $q$.

We now examine the condition of thermodynamic stability, $H(c,q)\leq 0$.
As it was already observed, since at the point $(c,q_0(c))$ both the
gradient of $\chi(c,q)$ and its Hessian determinant vanish, we have
$H\l(c,q_0(c)\r)=\l(\p^2 f/ \p c^2\r)\l(\p^2 \chi/\p q^2\r)\leq 0$. At the
theta point this quantity vanishes, and $H(c_\th,q)$ is given by the
deviations from the annealed approximation. Three situations are possible:
First, $H(c_\th,q)$ can be positive at the leading order in
$\d q=q-q_0(c_\th)$. In this case the thermodynamic stability would be
violated around the theta point, but our simulations do not show anything
strange in this region. Second, the leading order in $\d q$ can be
negative. In this case, the specific heat would not diverge anymore at the
theta point for finite $\D$, but it would show a peak proportional to some
negative power of $\d q$. Thus the disorder would smear out the thermodynamic
singularity at $c=c_\th$, leaving unchanged the geometric
characterization of the collapsed chains in terms of the gyration radius.
It is rather difficult, if not impossible, to
test this scenario by means of simulations. Third, $H(c_\th,q)$ can vanish
identically at $c=c_\th$. It is easy to see that this condition, combined
with the assumption that $\chi(c,q)$ is of the form (\ref{Plot}), is
fulfilled if and only if $\ti\chi^\pr(cq)$ is of the form

\be \ti\chi^\pr(cq)=\log\l({1-cq/B\over cq/A}\r)\: , \label{chi_pr}
\ee
where $q\geq q_0(c)$, $A<0$ and $0<B<cq$ are two constants, and $c$ is not
too small so that the last inequality can be fulfilled. In this case, one
finds $cq_0(c)=B(c-A)/(B-A)\in [0,c]$, and it is easy to check that all
previous results are recovered, while $H(c,q)-H(c,q_0(c))$ vanishes for all
$q$ and $c$, including the theta point.

Summarizing the discussion, we find that, if $\chi(c,q)$ is of the form
(\ref{Plot}), two possibilities are open: either $\ti\chi^\pr(cq)$ is given
by Eq.(\ref{chi_pr}), in which case $H(c,q)\equiv H(c,q_0(c))\leq 0$, or
$\ti\chi^\pr(cq)$ has a different form, in which case the specific heat
is not anymore divergent at the theta point $c=c_\th$. Unfortunately, we are
not able to decide among these alternatives.

%\bea {\p^2 Q_0\over \p c^2} & & =
%\l(a_2(c)\r)^{-2}\l(1-q_0(c)\r)^{-2}
%\l[\l(a_2(c)-ca'_2(c)\r)+ca_2(c)q'_0(c)\r] \nonumber \\
%&  & =-\l(1+cq_0(1-q_0){\p^2\ti\chi\over \p Q^2}\r)^{-3}
%\l(q_0^2(1-q_0)\r)\l[cq_0\l( \l({\p^2\ti\chi\over \p Q^2}\r)^2
%+ {\p^3\ti\chi\over \p Q^3} \r)+2{\p^2\ti\chi\over \p Q^2}
%\r].
%\eea
%
%If this term is positive, $H(c_\th,q)$ is negative.
%Although we do not know explicitly the function $\ti\chi(c,Q)$, it seems
%unlikely that the correction at $c=c_\th$ vanishes (while it would vanish
%under the hypothesis $\chi(c,q)=\ti\chi(cq)$). If it holds, as it should be
%for thermodynamic stability, $H(c_\th, q)<0$, this would mean that there is no
%thermodynamic singularity at the theta point for $\D>0$.
%In this case, while the geometric
%characterization of the collapsed chains in terms of the gyration radius
%would remain unchanged, the thermodynamics would change dramatically, smearing
%out the divergence of the specific heat at the theta point. Although the
%equations that we described suggest that this could be the case, we did
%not obtain from our numerical results a convincing indication of this fact.

\begin{figure}[ht]
  \centerline{
    \psfig{file=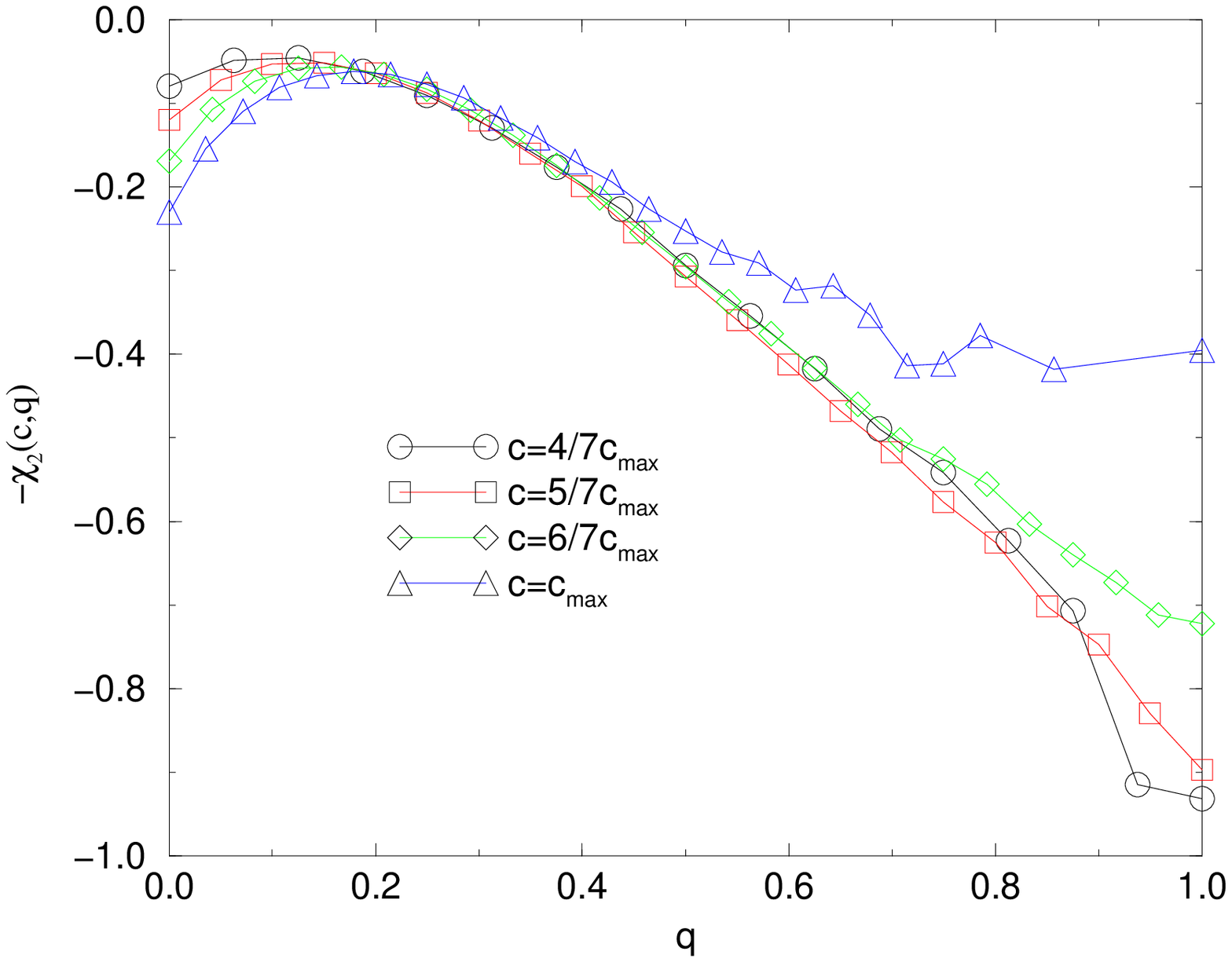,width=12cm}}
\caption{Entropy loss of a pair of chains with the same value of $c$ at
overlap $q$, computed as the logarithm of the probability density of the
overlap. Chains have $N=27$ monomers.}
\label{fig:chi27}
\end{figure}

We conclude this section showing in Fig.\ref{fig:chi27}, for the sake of
completeness, the
entropy loss $\chi_2(c,q)$ obtained from simulations of homopolymer
chains with $N=27$. Although the system size considered is definitely
too small for quantitative considerations, the qualitative behavior is
interesting and confirms the above considerations.
The entropy $\chi_2(c,q)$ was computed from

\be \chi_2(c,q)=-{1\over N}\log p_2(c,q), \ee
where
$p_2(c,q)$ is the probability density for a pair of chains, both with
density of contacts $c$, being at overlap $q$.

\section{Discussion}

We have shown that the annealed approximation is very good 
but not exact for a particular model of random heteropolymers, and we 
have given simple physical arguments for it.
We have also computed the thermodynamics of the model using the replica
symmetric approximation, and we have shown that such approach can explain
very well, at least qualitatively, the observed deviations from the annealed
approximation in the high temperature phase. The replica symmetric calculation
also leaves open, surprisingly, the possibility that the disorder could
cancel the thermodynamic singularity at the theta point. A numerical test of
this possibility is very difficult, and it has been left out.

In our present study we have not addressed the most interesting aspect 
of the model -- the freezing of the system in a finite number of mesoscopic
states.
This transition should represent some features of the folding transition
taking place for protein structures.
Instead, we
have studied the model at higher temperatures and at smaller disorder. 
This should however be of interest also in the context of the freezing,
since it was conjectured 
\cite{SG} that freezing can be described in this model by the 
random energy model, a prerequisite for which is that the annealed
approximation is exact.

In the present simulations we have studied chains of length up to $N=1400$. 
Deviations from the annealed approximation decrease fast for small $N$, 
which explains why studying very short chains has mislead several 
authors to the conclusion that these deviations vanish for $N\to\infty$.
But in the high-T (open coil) phase this decrease clearly stops, and
deviations are roughly
independent of $N$ for $N>100$. This is less clear in the collapsed phase. 
But also there numerics, general arguments, and detailed calculations 
within a specific scenario with unbroken replica symmetry all indicate 
that these deviations
will settle at a non-zero value for large $N$. This casts doubts
on the validity of the random energy picture for protein folding.

There are of course a number of questions which are left open by the 
present study. First of all, we have deliberately left out all questions 
related to freezing. Secondly, our treatment of sec.3 assumed that the 
overlap distribution is always dominated by a single peak. This is 
most likely not true in the frozen regime, and the distribution of 
overlaps is certainly a most interesting object. We shall address 
these questions in a forthcoming paper \cite{frozen}. Finally, we have 
studied only one particular model, where contact energies are independent 
Gaussian variables. Several other models of random heteropolymers 
are studied in the recent literature \cite{GOP}, and several of them 
present very intersting open problems.

\bigskip

We are indebted to many colleges for useful discussions, in particular 
to H. Frauenkron, W. Nadler, H. Orland, E. Shakhnovich, A. Trovato and
M. Vendruscolo.

\section*{Appendix: The Pruned Enriched Rosenbluth Method}

This method which was first described in detail in \cite{PERM} is a 
chain growth method. It is based on Rosenbluth's idea of biased 
sampling \cite{rosenbluth}, but it deviates from it by deleting (`pruning')
configurations with too low weight, and copying configurations with 
too large weight (`enrichment'). We remind the reader that the bias 
in the Rosenbluth Method requires each configuration to carry a 
non-trivial weight which builds up gradually as monomer by monomer is 
added. In addition to this `Rosenbluth factor', there is also a 
Boltzmann factor in the case of interacting polymers. The weights 
which control pruning/enrichment are the product of the two.

Both pruning and enrichment are done while the chains grow, i.e. 
based on the actual (incomplete) weight factors. In some cases it is 
advantageous to use not the present weights but (implicit) estimates 
of future weights to control the algorithm \cite{tube}, but this is 
not done in the present paper. Instead, we use some of the special 
tricks used for strongly collapsed systems in \cite{prot}. 

When the weight is too low, configurations are not simply killed 
(this would imply systematic errors), but instead they are killed 
with probability 1/2, and those which are not killed get their weights 
doubled. Similarly, in the case of cloning the weight is spread 
uniformly among all clones. Technically, cloning is performed by 
means of recursive subroutine calls. A pseudocode of the basic 
algorithm is given in the appendix of \cite{PERM}.

The most important shortcoming of the Rosenbluth Method is that 
the distribution of weights can become extremely wide for large 
systems and at low temperatures. The only exception is for 
interacting homopolymers on the simple cubic lattice at the theta 
point, where Rosenbluth and Boltzmann factors nearly cancel. In 
less favorable cases even a very large statistical sample can 
be dominated by just a handful of high weight events, and statistical 
errors grow out of bounds. Even worse, in extreme cases the events 
which would carry (in average!) most of the weight are so rare that 
they are missed completely with high probability, and the free energy 
is underestimated systematically.

Pruning and enrichment guarantee that the weights of individual 
configurations stay within narrow bounds, and the above cannot happen. 
But in very difficult situations (large $N$, low T, large disorder) 
it may happen that due to cloning the configurations are strongly 
correlated, and the weights of clusters of such correlated configurations 
play essentially a similar role as the weights of individual 
configurations in the above discussion. In the following we will call 
such a cluster which is composed of all configurations having a common 
root a `tour'. In order to check that the weights of tours do not become 
too uneven, we have measured in their distribution. Let us 
call the weights $W$, and the distribution $P(W)$. We are 
on safe grounds if this distribution is so narrow that $P(W)$ and 
$WP(W)$ have basically the same support. In particular, we should 
require that the maximum of $WP(W)$ occurs at such values where $P(W)$ 
is still appreciable, and the distribution is well sampled. We have verified 
that this is the case for all data shown in the present paper. Notice, 
however, that this is a very stringent requirement. If it is were not 
satisfied, this would not necessarily mean that the data are wrong, since 
configurations within one tour are only partially correlated. 

The generalization of this algorithm to pairs of chains growing 
simultaneously is straightforward \cite{CCG}. One just has to add 
monomers alternatingly. When the total number of monomers in both 
chains is even, the addition of the next monomer is attempted at chain
1; when this number is odd, the next monomer is added to chain 2.

\end{document}